# Dirac bound states of anharmonic oscillator in external fields

Majid Hamzavi[**,1], Sameer M. Ikhdair [*,2] and Babatunde J. Falaye [3,†]

[1] *Department of Physics, University of Zanjan, Zanjan, Iran*

[2] *Department of Physics, Faculty of Science, an-Najah National University, Nablus, West Bank, Palestine*

[3] *Theoretical Physics Section, Department of Physics, University of Ilorin, P. M. B. 1515, Ilorin, Nigeria.*

[**]Corresponding author Email: majid.hamzavi@gmail.com

[*]Email: sikhdair@gmail.com, sameer.ikhdair@najah.edu

[†] Email: fbjames11@physicist.net

## Abstract

We explore the effect of the external magnetic and Aharonov-Bohm (AB) flux fields on the energy levels of Dirac particle subjects to mixed scalar and vector anharmonic oscillator field in the two-dimensional (2D) space. We calculate the exact energy eigenvalues and the corresponding un-normalized two-spinor-components wave functions in terms of the chemical potential parameter, magnetic field strength, AB flux field and magnetic quantum number by using the Nikiforov-Uvarov (NU) method.

## 1. Introduction

Relativistic wave equations like the Dirac and the Klein–Gordon (KG) equations have received much interest from many theoretical physicists in various fields of physics [1,2]. Recently, the spin and pseudospin symmetries were predicted in the context of the Dirac equation. Within the framework of the Dirac equation, the pseudospin symmetry used to feature deformed nuclei and the superdeformation to establish an effective shell-model [3-8]. However, the spin symmetry is relevant to mesons [9] and occurs when the difference of the scalar $S(r)$ and vector $V(r)$ potentials are constant, i.e., $\Delta(r) = C_s$ while in the pseudospin symmetry it occurs when the sum of the scalar and vector potentials are constant, i.e., $\Sigma(r) = C_{ps}$ [10-13]. The exact pseudospin symmetry discussed in relativistic harmonic oscillator potential and a Woods-Saxon potential by Chen et al. [14]. A recent overview on the pseudo spin symmetry can be found in Ref. [15], and the readers are referred to Refs. [16-24] for some recent progress in this field.

In addition, over the past years, there has been an increased interest in finding exact bound state solutions to the relativistic equations with various vector and scalar potential models [25-30]. The most commonly used techniques in solving these wave equations are the Nikiforov–Uvarov (NU) method [31,32], supersymmetric quantum mechanics (SUSYQM) method [33,34], the point canonical transformation [35,36], the asymptotic iteration method [37,38], the exact quantization rule [39,40], the shifted N expansion technique (SNET) [41,42], the perturbative treatment approach [43] and the analytical exact iteration method (AEIM) [44].

The anharmonic oscillator potential which simply takes the form [45]

$$V(r) = ar^2 + \frac{b}{r^2}, \tag{1}$$

is extensively used to describe the bound states of the interaction systems and has been applied for both classical and modern physics. It plays a basic role in chemical and molecular physics as well since it can be used to calculate the molecular vibration–rotation energy spectrum of linear and non-linear systems [46-49]. In the non-relativistic quantum mechanics, the anharmonic oscillator potential is one of the exactly solvable potential models in the framework of the Schrödinger equation and has also been studied in one dimensional (1D), two dimensional (2D), three dimensional (3D), and even higher-dimensional space.

Very recently, we have studied the exact analytical bound state energy eigenvalues and normalized wave functions of the spinless relativistic equation with equal scalar and vector anharmonic oscillator potential interaction under the effect of external uniform magnetic field and AB flux field [50] by means of the Nikiforov–Uvarov (NU) method. The non-relativistic limit of our solution is obtained by making an appropriate mapping of parameters. Further, the KG- anharmonic and KG-harmonic oscillator special cases are also discussed. Furthermore, we carried out detailed exact energetic spectrum and wave functions of the Schrödinger equation with a anharmonic oscillator potential in the presence of external magnetic and AB flux fields [51,52]. Also the wave ansatz method is used to study of relativistic and nonrelativistic two-dimensional Cornell and Killingbeck potentials [53-54]

The problem of fermions in two dimensions interacting with classical background potentials has been an active area of research in Theoretical Physics. For instance, in quantum field theory, it has been used to describe some properties of quasi-one-dimensional conductors and some polymers [55]. It was explained in Ref. [55] that one of the interesting properties is the possibility of induced fractional fermion number in the vacuum due to interactions of fermions with topological backgrounds. To illustrate this phenomenon, one usually resorts to models of fermions interacting with background potentials of scalar and/or pseudoscalar natures.

The priority purpose of this work is therefore to solve the radial Dirac equation in two-dimensional (2D) space with the anharmonic oscillator potential in the presence of external constant magnetic and singular Aharonov-Bohm (AB) flux fields via the NU method. This method demands a trial ansatz for the wave function and is general enough to be applicable to a class of power and inverse power potential models.

The structure of this paper is as follows. In Section 2, we briefly introduce the solution of the Dirac equation with scalar-vector potentials in the presence of external magnetic and AB flux fields. Further, the radial Dirac equation is applied to deal with scalar-vector anharmonic oscillator potentials. The outlines of the Nikiforov-Uvarov method are presented in Section 3. The energy eigenvalue equations and the corresponding eigenfunctions are obtained in Section 4. Some numerical results are also given. We end with conclusion in Section 5.

## 2. Dirac Equation with Scalar-Vector Potentials

The Dirac equation for fermionic massive spin-$1/2$ particles moving in an attractive scalar potential $S(r)$ and a repulsive vector potential $V(r)$ (in units of $\hbar = c = 1$) is

$$\left[\vec{\alpha}.\vec{p} + \beta(M + S(r))\right]\psi(\vec{r}) = \left[E - V(r)\right]\psi(\vec{r}), \tag{2}$$

where $E$ is the relativistic energy of the system, $\vec{p} = -i\vec{\nabla}$ is the three-dimensional momentum operator and $M$ is the fermionic particle mass. $\vec{\alpha}$ and $\beta$ are the $4 \times 4$ usual Dirac matrices give as

$$\vec{\alpha} = \begin{pmatrix} 0 & \vec{\sigma} \\ \vec{\sigma} & 0 \end{pmatrix}, \quad \beta = \begin{pmatrix} I & 0 \\ 0 & -I \end{pmatrix}, \tag{3}$$

where $I$ is $2 \times 2$ unitary matrix and $\vec{\sigma}$ are three-vector spin matrices

$$\sigma_1 = \begin{pmatrix} 0 & 1 \\ 1 & 0 \end{pmatrix}, \quad \sigma_2 = \begin{pmatrix} 0 & -i \\ i & 0 \end{pmatrix}, \quad \sigma_3 = \begin{pmatrix} 1 & 0 \\ 0 & -1 \end{pmatrix}. \tag{4}$$

In the Pauli–Dirac representation,

$$\psi(\vec{r}) = \begin{pmatrix} \varphi(\vec{r}) \\ \chi(\vec{r}) \end{pmatrix}, \tag{5}$$

where $\varphi(\vec{r})$ is the upper (large) component and $\chi(\vec{r})$ is the lower (small) component of the Dirac spinors, and substituting Eq. (5) into Eq. (2), one obtains two coupled differential equations for upper and lower radial wave functions as

$$\vec{\sigma}.\vec{p}\chi(\vec{r}) = [E - M - \Sigma(r)]\varphi(\vec{r}), \quad (6a)$$

$$\vec{\sigma}.\vec{p}\varphi(\vec{r}) = [E + M - \Delta(r)]\chi(\vec{r}), \quad (6b)$$

where we have introduced the difference and sum of potentials as

$$\Delta(r) = V(r) - S(r), \quad (7a)$$

$$\Sigma(r) = V(r) + S(r), \quad (7b)$$

respectively. In what follows we shall investigate the spin and pseudospin symmetries in the Dirac Hamiltonians.

## 2.1. Exact Pseudospin Symmetric Limit $(\Sigma(r) = 0)$

Ginocchio showed that there is a connection between pseudospin symmetry and near equality of the time component of a vector potential and the scalar potential, $V(r) = -S(r)$. After that, Meng et al. showed that if $\frac{d[V(r)+S(r)]}{dr} = \frac{d\Sigma(r)}{dr} = 0$ or $\Sigma(r) = C_{ps} =$ constant, then pseudospin symmetry is exact in the Dirac equation [56-62]. In this part, we are taking the difference potential $\Delta(r)$ as anharmonic oscillator potential. Thus, Eqs. (6a) and (6b) become

$$\vec{\sigma}.\vec{p}\varphi(\vec{r}) = [E + M - 2V(r)]\chi(\vec{r}), \quad (8a)$$

$$\varphi(\vec{r}) = \frac{\vec{\sigma}.\vec{p}}{E - M}\chi(\vec{r}). \quad (8b)$$

Therefore under this symmetry and combining the two equations in (8), one obtains

$$\left[\vec{p}^2 + 2(E-M)V(r)\right]\chi(\vec{r}) = \left(E^2 - M^2\right)\chi(\vec{r}). \quad (9)$$

Now to study the effect of external magnetic field and AB flux field on a spin-1/2 fermionic particle exposed to a anharmonic oscillator field, we make a simple transformation $\vec{p} \rightarrow \vec{p} - \frac{e}{c}\vec{A}.$ Hence, the Dirac equation for a charged particle moving in a constant magnetic and AB flux fields becomes

$$\left[\left(\vec{p}-\frac{e}{c}\vec{A}\right)^2 + 2(E-M)V(r)\right]\chi(\vec{r}) = \left(E^2 - M^2\right)\chi(\vec{r}), \qquad (10)$$

with the vector potentials [44,50-54]

$$\vec{A} = \vec{A}_1 + \vec{A}_2,$$

$$\vec{A}_1 = \frac{1}{2}\vec{B}_0 \times \vec{r} = \frac{B_0 r}{2}\hat{\phi}, \quad \vec{A}_2 = \frac{\Phi_{AB}}{2\pi r}\hat{\phi}, \qquad (11)$$

such that $\vec{\nabla}\times\vec{A}_1 = B_0\hat{k}$ and $\vec{\nabla}\times\vec{A}_2 = 0$ where $\vec{B}_0$ is the applied magnetic field that is perpendicular to the plane $(r,\phi)$ and $\vec{A}_2$ describes the additional magnetic flux $\Phi_{AB}$ created by a solenoid. Further, $S(\vec{r})$ and $V(\vec{r})$ are the scalar and vector potentials, respectively. Here, we shall solve Eq. (10) with unequal mixture of scalar-vector anharmonic oscillator potentials (1), i.e., $S(r) = -V(r)$, by using the 2D cylindrical trial wave function given by

$$\chi_{nm}(\vec{r}) = \frac{1}{\sqrt{r}} g_{nm}(r) e^{im\phi}, \quad m = 0, \pm 1, \pm 2, \cdots. \qquad (12)$$

where $m$ is a new value of angular momentum (magnetic quantum number). By the above substitution we hereby obtain a second order radial differential equation satisfying the radial part of wave function $g_{nm}(r)$ as

$$\frac{d^2 g_{nm}(r)}{dr^2} - \left[ 2(E-M)ar^2 + \frac{e^2 B^2 r^2}{4c^2} + 2(E-M)\frac{b}{r^2} \right.$$
$$\left. + \frac{(m'^2 - 1/4)}{r^2} + \frac{e^2 B \phi_{AB}}{2\pi c^2} - \frac{emB}{2c} - \left(E^2 - M^2\right) \right] g_{nm}(r) = 0, \qquad (13)$$

where $m'^2 = m^2 + \frac{e^2 \phi_{AB}^2}{4\pi^2 c^2} - \frac{em\phi_{AB}}{\pi c}$.

**2.2. Exact Spin Symmetric Limit ($\Delta(r) = 0$)**

In the spin symmetric limit $\frac{d\Delta(r)}{dr} = 0$ or $\Delta(r) = C_s = \text{constant}$ [56-60], then Eq. (6) becomes

$$\vec{\sigma}.\vec{p}\chi(\vec{r}) = [E - M - 2V(r)]\varphi(\vec{r}), \tag{14a}$$

$$\chi(\vec{r}) = \frac{\vec{\sigma}.\vec{p}}{E + M}\varphi(\vec{r}). \tag{14b}$$

Further, combining Eqs. (14a) and (14b), we obtain

$$\left[\vec{p}^2 + 2(E + M)V(r)\right]\varphi(\vec{r}) = \left(E^2 - M^2\right)\varphi(\vec{r}). \tag{15}$$

Therefore, the Dirac equation for a charged particle moving in a constant magnetic and AB flux fields appears as

$$\left[\left(\vec{p} - \frac{e}{c}\vec{A}\right)^2 + 2(E + M)V(r)\right]\varphi(\vec{r}) = \left(E^2 - M^2\right)\varphi(\vec{r}), \tag{16}$$

with $\vec{A}$ is defined as before (see Eq. (11)) and setting the upper spinor wavefunction as

$$\varphi_{nm}(\vec{r}) = \frac{1}{\sqrt{r}} f_{nm}(r) e^{im\phi}, \quad m = 0, \pm 1, \pm 2, \cdots. \tag{17}$$

Substituting Eqs. (11) and (17) into Eq. (16) and with a little algebra we can obtain the following Schrödinger-like equation:

$$\frac{d^2 f_{nm}(r)}{dr^2} - \left[ 2(E+M)ar^2 + \frac{e^2 B^2 r^2}{4c^2} + 2(E+M)\frac{b}{r^2} \right.$$
$$\left. + \frac{(m'^2 - 1/4)}{r^2} + \frac{e^2 B \phi_{AB}}{2\pi c^2} - \frac{emB}{2c} - \left(E^2 - M^2\right) \right] f_{nm}(r) = 0. \tag{18}$$

Now, Eqs. (13) and (18) can be solved in the framework of the Nikiforov-Uvarov method which shall be briefly introduced in the following section.

## 3. The Method of Analysis

The Nikiforov-Uvarov method [63] is based on solving the hypergeometric type second-order differential equations by means of special orthogonal functions. The main equation which is closely associated with the method is given in the following form

$$\psi_n''(s) + \frac{\tilde{\tau}(s)}{\sigma(s)}\psi_n'(s) + \frac{\tilde{\sigma}(s)}{\sigma^2(s)}\psi_n(s) = 0, \tag{19}$$

where $\sigma(s)$ and $\tilde{\sigma}(s)$ are polynomials, at most of second-degree, and $\tilde{\tau}(s)$ is a first-degree polynomial. Let us discuss the exact particular solution of Eq. (19) by choosing

$$\psi_n(s) = y_n(s)\phi(s), \tag{20}$$

resulting in a hypergeometric type equation of the form:

$$\sigma(s)y_n''(s) + \tau(s)y_n'(s) + \lambda y_n(s) = 0, \tag{21}$$

where $\lambda$ is a constant given in the form

$$\lambda = \lambda_n = -n\tau'(s) - \frac{n(n-1)}{2}\sigma''(s), \quad n = 0, 1, 2, \cdots. \tag{22}$$

The first part of the wave function, i.e. $y_n(s)$, is the hypergeometric-type function whose polynomials solutions are given by Rodrigue's relation

$$y_n(s) = \frac{B_n}{\rho(s)} \frac{d^n}{ds^n}\left[\sigma^n(s)\rho(s)\right], \tag{23}$$

where $B_n$ is the normalization constant and the weight function $\rho(s)$ must be satisfied by the condition

$$\frac{d}{ds}\left[\sigma(s)\rho(s)\right] = \tau(s)\rho(s), \tag{24}$$

$$\tau(s) = \tilde{\tau}(s) + 2\pi(s), \tag{25}$$

and the derivative of $\tau(s)$ must be negative [35]. The function $\pi(s)$ and the parameter $\lambda$ required for this method are defined as follows

$$\pi(s) = \frac{\sigma' - \tilde{\tau}}{2} \pm \sqrt{\left(\frac{\sigma' - \tilde{\tau}}{2}\right)^2 - \tilde{\sigma} + k\sigma}, \tag{26}$$

$$\lambda = k + \pi'(s). \tag{27}$$

In order to find the value of $k$, the expression under the square root must be square of polynomial. At the end, the second part of wave function in Eq. (20), i.e. $\phi(s)$, is defined as logarithmic derivative

$$\frac{\phi'(s)}{\phi(s)} = \frac{\pi(s)}{\sigma(s)}. \tag{28}$$

## 4. Bound State Solutions to Dirac Equation

We are going to solve the Dirac equation with anharmonic oscillator potential under spin and pseudospin symmetries by using the Nikiforov-Uvarov method.

### 4.1. Pseudospin symmetric case

Now, we seek to solve Eq. (13) with the following change of variables $s = r^2$ and obtain

$$\left[ \frac{d^2}{ds^2} + \frac{1}{2s}\frac{d}{ds} - \frac{1}{4s^2}\left( \tilde{p}^2 s^2 + \tilde{q} s + \tilde{\delta} \right) \right] g_{nm}(s) = 0, \tag{29}$$

where

$$\begin{aligned}
\tilde{p}^2 &= 2(E-M)a - \frac{e^2 B^2}{4c^2}, \\
\tilde{q} &= \frac{e^2 B \phi_{AB}}{2\pi c^2} - \frac{emB}{2c} - \left( E^2 - M^2 \right), \\
\tilde{\delta} &= m^2 + \frac{e^2 \phi_{AB}^2}{4\pi^2 c^2} - \frac{em\phi_{AB}}{\pi c} - \frac{1}{4} - 2(E-M)b.
\end{aligned} \tag{30}$$

Further, we compare Eq. (29) with Eq. (19) to obtain the following particular values for the parameters:

$$\begin{aligned}
\tilde{\tau}(s) &= 1, \\
\sigma(s) &= 2s, \\
\tilde{\sigma}(s) &= -\left( \tilde{p}^2 s^2 + \tilde{q} s + \tilde{\delta} \right),
\end{aligned} \tag{31}$$

and making use of Eq. (26), $\pi(s)$ can be easily obtained as

$$\pi(s) = \begin{cases} \frac{1}{2} \pm \left( \tilde{p} s + \sqrt{\tilde{\delta} + \frac{1}{4}} \right) & \text{for} \quad k = -\frac{\tilde{q}}{2} + \frac{\tilde{p}}{2}\sqrt{4\tilde{\delta} + 1} \\ \frac{1}{2} \pm \left( \tilde{p} s - \sqrt{\tilde{\delta} + \frac{1}{4}} \right) & \text{for} \quad k = -\frac{\tilde{q}}{2} - \frac{\tilde{p}}{2}\sqrt{4\tilde{\delta} + 1} \end{cases}. \tag{32}$$

The essential point here is to have the derivative of $\tau(s)$ be negative so that one obtains bound state solutions. Thus, we must have the proper choice:

$$\pi(s) = \frac{1}{2} \pm \left( \tilde{p} s - \sqrt{\tilde{\delta} + \frac{1}{4}} \right) \quad \text{for} \quad k = -\frac{\tilde{q}}{2} - \frac{\tilde{p}}{2}\sqrt{4\tilde{\delta} + 1}. \tag{33}$$

Therefore, from Eq. (25) $\tau(s)$ takes the form:

$$\tau(s) = 2 - 2\tilde{p} s + \sqrt{\tilde{\delta} + \frac{1}{4}}, \tag{34}$$

and from Eqs. (22) and (27), one obtains

$$\lambda = \lambda_n = -n\tau'(s) - \frac{n(n-1)}{2}\sigma''(s) = 2\tilde{p} n, \tag{35}$$

and

$$\lambda = k + \pi'(s) = -\frac{\tilde{q}}{2} - \frac{\tilde{p}}{2}\sqrt{4\tilde{\delta}+1} - \tilde{p}. \tag{36}$$

The energy eigenvalue equation can be found by equating Eqs. (35) and (36) as follows

$$\tilde{p}\left(4n+2+\sqrt{4\tilde{\delta}+1}\right) + \tilde{q} = 0, \tag{37}$$

and making use of Eq. (30), we finally obtain the energy equation:

$$2\sqrt{2(E-M)a + \frac{e^2 B^2}{4c^2}} \left( 2n+1 + \sqrt{m^2 + \frac{e^2 \phi_{AB}^2}{4\pi^2 c^2} - \frac{em\phi_{AB}}{2\pi c} + 2(E-M)} \right)$$
$$+ \frac{eB\phi_{AB}}{2\pi c^2} - \frac{emB}{2c} - (E^2 - M^2) = 0 \tag{38}$$

Our numerical results of the energy states in pseudospin symmetric case are plotted against the magnetic and AB flux fields in Figures 1 and 2, respectively. In the presence of exact pseudospin symmetry, it is noticed in Figure 1 that the energy levels increase with the increasing magnetic field strength $B$ for constant values of principal $n$ and magnetic $m$ quantum numbers when the AB flux field $\phi_{AB}$ has a constant strength. At large value of $B$, the energy becomes linear in shape while at small values it appears parabolic. Further, the energy fan out toward the increasing energy when the magnetic quantum numbers $m$ increase for a given $n$ state as the magnetic quantum number is mainly dependent on $B$ [50]. The magnetic field splits energy states up/down according to the values of $m$. In Figure 2, the energy overlaps at some values of AB flux field $\phi_{AB}$ with a constant value of $B$ for non zero magnetic quantum numbers $m \neq 0$. However, there is no overlapping when $m=0$.

Next to find eigenfunction of pseudospin symmetry case, we first determine the weight function from Eqs. (24), (31) and (34) as

$$\rho(s) = e^{-\tilde{p}s} s^{\sqrt{\tilde{\delta}+1/4}} \tag{39}$$

then, by substituting Eq. (39) into Eq. (23), we find the first part of wave function as

$$y_n(s) = \frac{B_n}{\rho(s)} \frac{d^n}{ds^n} \left[ \sigma^n(s) \rho(s) \right]$$
$$= \frac{B_n}{e^{-\tilde{p}s} s^{\sqrt{\tilde{\delta}+1/4}}} \frac{d^n}{ds^n} \left[ (2s)^n e^{-\tilde{p}s} s^{\sqrt{\tilde{\delta}+1/4}} \right] = B_n L_n^{\sqrt{\tilde{\delta}+1/4}}(\tilde{p}s), \tag{40}$$

where $B_n$ is normalization constant and $L_n^k(s)$ is the generalized Laguerre polynomial. For second part of wave function, i.e. $\phi(s)$, we use Eq. (28) and we obtain

$$\phi(s) = e^{-\frac{\tilde{p}s}{2}} s^{\frac{1}{4}(1+\sqrt{4\tilde{\delta}+1})}, \tag{41}$$

Finally, the lower radial component of the Dirac spinor is obtained as ($s = r^2$)

$$g_{nm}(r) = \phi(r)y_n(r) = \tilde{D}_{nm} e^{-\tilde{p}r^2/2} r^{(1+\sqrt{4\tilde{\delta}+1})/2} L_n^{\sqrt{\tilde{\delta}+1/4}}(\tilde{p}r^2), \tag{42}$$

which satisfies the asymptotic behavior at $r = 0$ and $r \to \infty$.

### 4.2. Spin symmetric case

To obtain a solution of Eq. (18) and to avoid repetition, we follow similar procedure like the ones in subsection 4.1. We find the following relation for energy eigenvalues as

$$p\left(4n + 2 + \sqrt{4\delta + 1}\right) + q = 0, \tag{43}$$

which gives the energy equation in the presence of spin symmetry as

$$2\sqrt{2(E+M)a + \frac{e^2 B^2}{4c^2}} \left(2n + 1 + \sqrt{m^2 + \frac{e^2 \phi_{AB}^2}{4\pi^2 c^2} - \frac{em\phi_{AB}}{2\pi c} + 2(E+M)}\right) \\
+ \frac{eB\phi_{AB}}{2\pi c^2} - \frac{emB}{2c} - (E^2 - M^2) = 0 \tag{44}$$

and the corresponding wave functions for upper the Dirac spinor as

$$f_{nm} = D_{nm} e^{-pr^2/2} r^{(1+\sqrt{4\delta+1})/2} L_n^{\sqrt{\delta+1/4}}(pr^2), \tag{45}$$

where

$$p^2 = 2(E+M)a - \frac{e^2 B^2}{4c^2},$$
$$q = \frac{eB\phi_{AB}}{2\pi c^2} - \frac{emB}{2c} - \left(E^2 - M^2\right), \tag{46}$$
$$\delta = m^2 + \frac{e^2 \phi_{AB}^2}{4\pi^2 c^2} - \frac{em\phi_{AB}}{\pi c} - \frac{1}{4} - 2(E+M)b.$$

To see the behavior of energy states in the presence of different fields, we plot the numerical results of the energy states in spin symmetric case against the magnetic and AB flux fields in Figures 3 and 4, respectively.

In the exact spin symmetry, the energy in Figure 3 is slightly increasing for $m \neq 0$ states when the magnetic field increases with $\phi_{AB}$ is constant in magnitude; this means that splitting of energy states becomes narrower when $B$ gets stronger. In Figure 4, the effect of change in $\phi_{AB}$ on the energy is that they overlap at large values of $\phi_{AB}$.

## 5. Concluding Remarks

In this work, we investigated the exact energy levels and corresponding wave function for arbitrary $m$-state in two-dimensional Dirac equation with scalar and vector anharmonic oscillator potentials under the influence of external uniform magnetic and singular AB flux fields in the context of analytical NU method. We studied the energy equations and wave functions under spin and pseudospin symmetries. We have also shown the effect of magnetic quantum numbers $m \neq 0$ on the energy states due to the existence of magnetic field $B$. This magnetic field produces a splitting in the energy states. However, the AB flux field makes the energy states overlapping for $m \neq 0$ states at large strength of $\phi_{AB}$. It is found that the energy spectrum is increasing with the magnetic quantum numbers $m$.

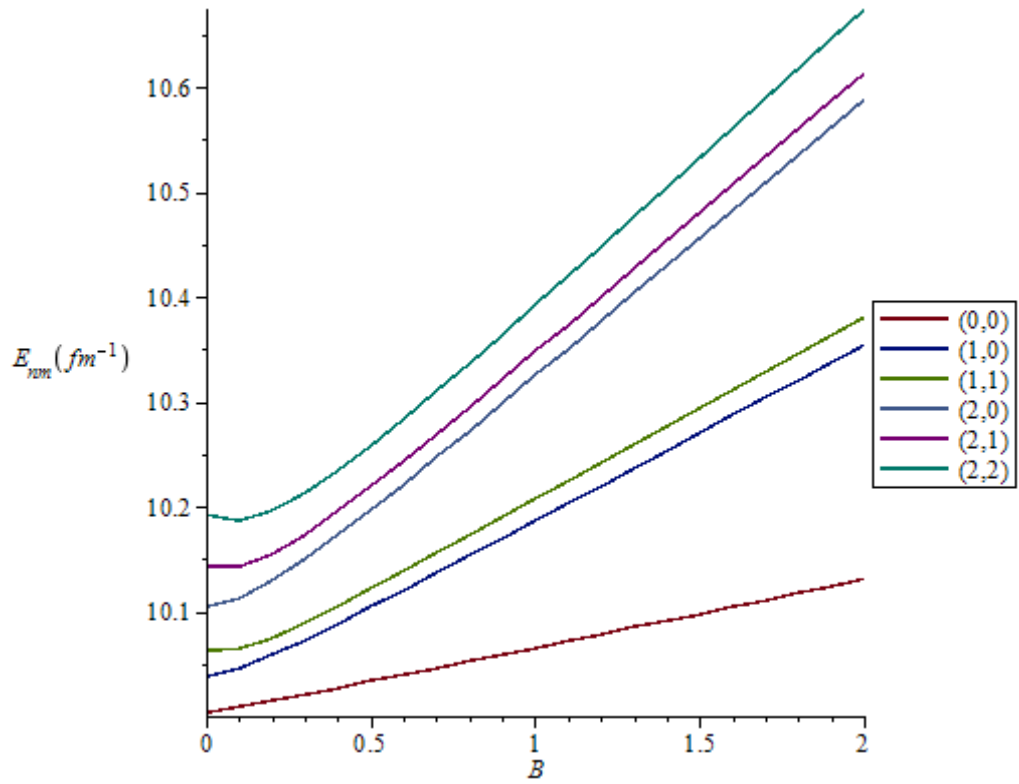

**Figure 1.** Pseudospin symmetric eigenenergies as a function of magnetic field.

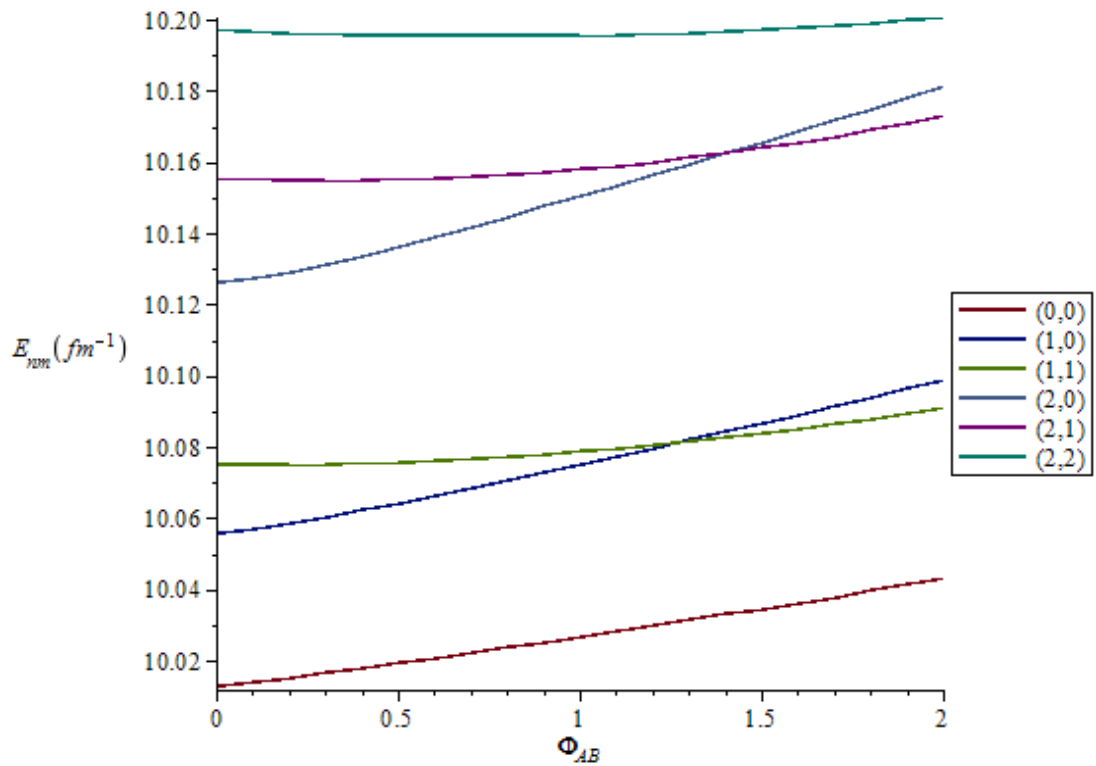

**Figure 2.** Pseudospin symmetric eigenenergies as a function of AB flux field.

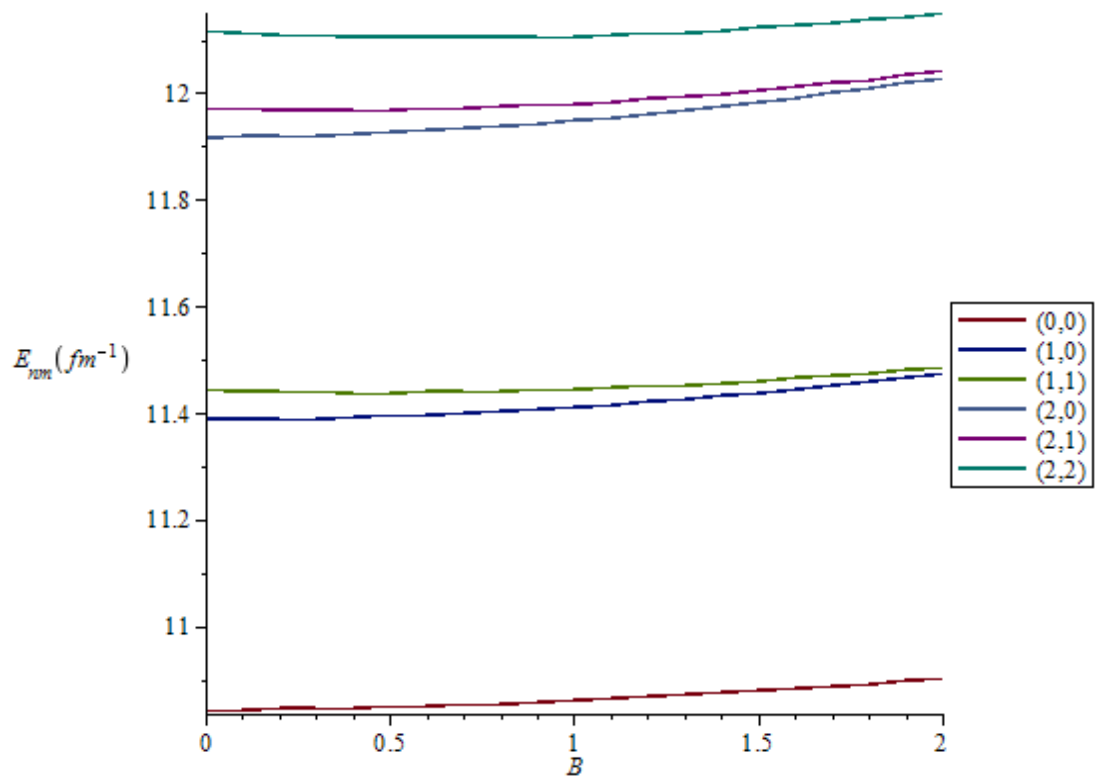

**Figure 3.** Spin symmetric eigenenergies as a function of magnetic field.

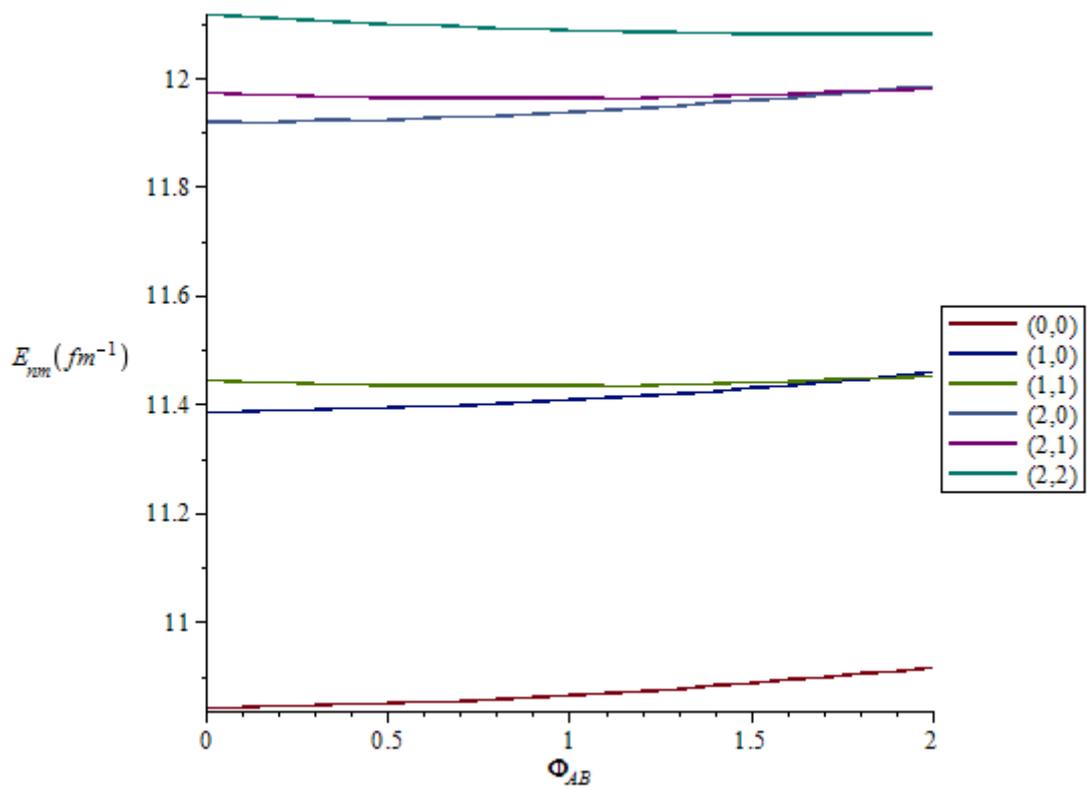

**Figure 4.** Spin symmetric eigenenergies as a function of AB flux field.